\documentstyle[12pt,aaspp4]{article}

\def\t0{\theta_{\circ}}

\def\be{\begin{equation}}
\def\en{\end{equation}}

\def\etal{et al.\ }
\def\msun{\rm M_{\odot}}

\def\msunyr{M_{\sun} yr^{-1}}
\def\kms{\rm \, km \, s^{-1}}
\def\mdot{\dot{M}}

\def\MSUNYR{\rm M_{\odot}\,yr^{-1}}
\def\MSUN{\rm M_{\odot}}

\def\RSUN{\rm R_{\odot}}
\def\Mdot{ \dot{M}}
\def\prom#1{\langle #1\rangle}
\def\cm2g{\rm cm^2 \ g^{-1}}

\begin{document}

\title
{Evidence for a developing gap in a 10 Myr old protoplanetary disk}

\author{  Nuria Calvet \altaffilmark{1}, Paola D'Alessio \altaffilmark{2,3}, Lee Hartmann\altaffilmark{1}, 
David Wilner\altaffilmark{1}, Andrew Walsh\altaffilmark{4}, 
and Michael Sitko\altaffilmark{5}} 

\altaffiltext{1}{Harvard-Smithsonian Center for Astrophysics, 60 Garden St., Cambridge, MA 02138, USA.
Electronic mail: ncalvet@cfa.harvard.edu, paola@amnh.org, lhartmann@cfa.harvard.edu, 
dwilner@cfa.harvard.edu, awalsh@mpifr-bonn.mpg.de, sitko@physics.uc.edu}
\altaffiltext{2}{Instituto de Astronomia, UNAM, Ap.P. 70-264, 04510 M\'exico D.F., M\'exico}
\altaffiltext{3}{American Museum of National History, Central Park West at 79th Street,
New York, NY10024-5192}
\altaffiltext{4}{Max Planck Institut f\"{u}r Radioastronomie, auf dem H\"{u}gel 69, Bonn,
53121, Germany}
\altaffiltext{5}{Department of Physics, University of Cincinnati, Cincinnati, OH 45221-0011}

\begin{abstract}

We have developed a physically self-consistent model of the
disk around the nearby 10 Myr old star TW Hya which 
matches the observed spectral energy distribution and
7mm images of the disk.  The model requires
both significant dust size evolution and a partially-evacuated
inner disk region, as predicted by theories of planet formation.
The outer disk, which extends to at least 140 AU in radius, is 
very optically thick at infrared wavelengths and quite massive 
($\sim 0.06 \msun$) for the relatively advanced
age of this T Tauri star.  This implies long viscous and dust 
evolution timescales, 
although dust must have grown to sizes
of order $\sim 1$~cm to explain the sub-mm and mm spectral slopes.
In contrast, the negligible near-infrared
excess emission of this system requires that the disk be optically 
thin inside $\lesssim$ 4 AU.  This inner region cannot be completely
evacuated; we need $\sim $ 0.5 lunar mass of $\sim$ 1 $\mu$m particles
remaining to produce the observed $10 \mu$m silicate emission.
Our model requires a distinct transition in disk properties
at $\sim 4$ AU, separating the inner and outer disk. 
The inner edge of the optically-thick outer disk must be heated almost frontally 
by the star to account for the 
excess flux at mid-infrared wavelengths.
We speculate that this truncation of the outer disk may be the signpost 
of a developing gap due to the effects of a growing protoplanet; the
gap is still presumably evolving because material still resides in
it, as indicated by the silicate emission, the molecular hydrogen emission,
and by the continued accretion onto the central star (albeit at a much
lower rate than typical of younger T Tauri stars).
TW Hya thus may become the Rosetta stone for our
understanding of the evolution and dissipation
of protoplanetary disks.

\end{abstract}

\keywords{Accretion, accretion disks, Stars: Circumstellar Matter,
Stars: Formation, Stars: Pre-Main Sequence}

\section{Introduction}

The discovery of extrasolar planets (Marcy \& Butler 1998 and references therein) has opened up a new era
in the study of planetary systems.  While many important clues
to the processes of planet formation can be obtained from studies
of older systems, the best tests of formation scenarios will require
the direct detection of actively planet-forming systems.

It is thought that the formation of giant planets involves the sweeping
up of material in a wide annulus in the circumstellar disk, resulting
in the development of a gap (Lin \& Papaloizou 1986, 1993; Bryden \etal 1999).
Material inside the planet-driven gap can continue to accrete 
onto the central star; if the planet can prevent material from accreting
across the gap into the inner disk, 
the eventual result would be the evacuation of the region 
interior to the planet.  In the case of the Solar System, 
the formation of Jupiter might have prevented outer disk gas from
reaching the inner solar system; the inner gas disk accreted into the
Sun, while solid planetesimals remaining behind eventually formed the
terrestrial planets.

The above scenario suggests that the signature of a forming giant planet
would be the presence of a gap that is not entirely evacuated.
In this case dusty emission from the inner disk might still be observable. 
In addition, giant planet formation requires the
prior consolidation of large solid  bodies to serve as cores for subsequent
gas accretion; while such bodies would be invisible with current
techniques, one would expect to see evidence for substantial growth in
dust particles.  Finally, if the inner disk has not been completely
evacuated by the forming planet, one might expect to observe continued 
accretion onto the central star, as all T Tauri systems with inner disks 
as detected from near-infrared disk emission are also accreting
(Hartigan \etal 1990).

In this article we 
propose 
that the relatively young low-mass star
TW Hya has a developing gap in its inner disk qualitatively similar to
that expected from planet formation.  
The evidence supporting this this proposal is:
(1) reduced (optically-thin) emission from the inner disk; (2) mm-wave spectra
which seem to require grain growth; (3) extra emission from the edge of the
outer disk; and (4) continued accretion onto the
central star, albeit at a rate substantially lower than that observed from
most T Tauri stars.  TW Hya has an age of 10 Myr and so according to
current theories it is quite likely to be close to the epoch of planet 
formation.  It is also part of an association of young stars of similar
age, but it stands out in that it is the only system still accreting
at a substantial rate (Muzerolle \etal 2000).  The (outer) disk of TW Hya 
is quite massive, so that there is likely to be more than enough material available
to form giant planet(s).  TW Hya is also the closest known such system, 
and so it will be a prime target for following studies to confirm our model.

\section{ Observations}

The observations used to constrain our disk model are taken from the literature.
In addition, we have obtained
a narrow-band L (3.55-3.63 microns) measurement
of TW Hya on June 14th, 2000 with the Australian National
University 2.3-m telescope at Siding Spring, using the near infrared
camera CASPIR (Cryogenic Array Spectrometer Imager; McGregor et al. 1994).
CASPIR contains a 256x256 InSb array. TW Hya was imaged
at L with a pixel scale of 0.25 arcseconds
and an on-source integration time of 236 seconds.
The standard star BS4638 (magnitude 4.50 at L) was observed at a similar
airmass to TW Hya and was used to calibrate the magnitude of TW Hya.
The L magnitude we obtained was 7.12, equivalent to 0.39Jy.

This measurement is important in constraining the amount of near-infrared
excess in the system. Using K=7.37 (Webb et al. 1999), K-L = 0.25.
The K-L color for a K7V star is 0.11 (Kenyon \& Hartmann 1995),
which would yield an infrared excess of 0.14. 
However, such small
infrared excess does not necessarily imply emission from
disk material (Wolk \& Walter 1996).
For example,
the non-accreting stars in Taurus 
(spectral types $\sim$ K7 to M2)
have K-L between -0.05 and 0.25,
while 95\% of the accreting stars in Taurus 
have K-L between 0.35 and 1.2
(Meyer, Calvet, \& Hillenbrand 1997). 
This small value of K-L is consistent with the simultaneous
flux measurements of Sitko \etal (2000) indicating little if any
hot dust emission at wavelengths $< 5 \mu$m, though it is difficult
to rule out completely any excess.

\section{Model assumptions}

We assume that the heating of the disk is due to stellar irradiation
and viscous dissipation, and calculate self-consistently
the disk heights and temperatures
following the methods 
we have applied to interpret young $\sim$ 1 Myr old
disks (D'Alessio et al. 1998 = Paper I, 1999 = Paper II, 2000 = Paper III.)
We adopt $M_*=0.6 \ \MSUN$, $R_*=1 \ \RSUN$ and $T_*=4000$ K
for the stellar mass, radius, and effective temperature 
(Webb et al. 1999), and a mass accretion rate of
$\Mdot=5 \times 10^{-10} \ \MSUNYR$ as derived by Muzerolle \etal (2000),
which we take as constant through the whole disk.
The inclination of the disk axis to the line of sight 
is $i \sim 0^{\circ}$, in agreement with the
nearly symmetric HST images (Krist et al. 2000), and we adopt
the Hipparcos distance of 55 pc (Wichmann et al. 1998).

With the assumption of steady accretion, the disk surface density
scales as $\Sigma \propto \Mdot \alpha^{-1} T^{-1}$, 
with $T$ is the midplane temperature (see Paper II). 
The temperature does not vary much between acceptable models, 
partly because it is mostly determined by irradiation heating,
not viscous energy dissipation.  The disk mass therefore is roughly
constant for constant $\Mdot \alpha^{-1}$. Moreover, for long
wavelengths where much of the disk is optically thin, the emergent fluxes 
also tend to scale in the same way. 
For the purposes of fitting, we   
vary the parameter $\alpha$ 
as the dust 
opacities are varied, but note that varying $\alpha$ is equivalent 
to varying the disk mass, because $\Mdot$ is fixed. To first order, 
however, choices with 
$\Mdot \alpha^{-1} \approx$~constant are acceptable.

We use a dust mixture consisting of silicates, refractory organics,
troilite and water ice, following Pollack et al. (1994 = P94).  The dust
size distribution is taken to be $n(a) \sim a^{-p}$, with $p = 3.5$, between given
$a_{min}$ and $a_{max}$.  Optical properties for the compounds are
taken from J\"ager et al. (1994), P94, Begemann et al. (1994; also see 
Henning et al. 1999), and Warren (1984).  
We consider the grains to be compact segregated spheres, and calculate 
the opacity using a Mie scattering code (Wiscombe 1979). 
Sublimation temperatures
for the different grain types are taken from P94.

\section{Disk model}

The spectral energy distribution (SED) of TW Hya is shown in Figure 1. 
References for the observational data are given in the
figure caption. Note that
our narrow-band L observation
overlaps with Sitko et al. (2000) fluxes.

In Figure 1, 
we compare the SED with the median SED for $\sim 1$ Myr-old
T Tauri stars in the Taurus molecular cloud (Paper II),
normalized to the TW Hya stellar photosphere
at $H$ ($1.6 \mu$m), thus compensating for the differing distances
and stellar luminosities.
Even though the fluxes of TW Hya are relatively high compared
with the median Taurus SED at $\lambda \gtrsim 20 \mu$m, there is
a large flux deficit in TW Hya below 10 microns
(Jayawardhana et al. 1999);
in particular, fluxes are essentially photospheric
below  $ \sim 6 \mu$m (Sitko et al. 2000). 
The flux deficit below 10 $\mu$m has lead to
inferences of disk clearing inside a few AU from the central
star (Jayawardhana et al. 1999), but 
since the disk is still accreting
mass onto the star (Muzerolle et al. 2000),
it has to extend all the way into the corotation
radius at least, so the inner disk radius has to be
$\lesssim$ 0.03 AU (inferred from the 2 day photometric
period [Mekkaden 1998]).

We show below that models with a uniform dust 
well-mixed with the gas
throughout the disk
extending all the way onto the inner radius
cannot explain these features of the SED. 
The observations are much better understood if
the disk is divided into two regions:
the outer disk, which is more nearly comparable to the structure
inferred for typical T Tauri disks (Paper III), and the inner disk, which
is much more optically thin than in typical T Tauri disk models.

\subsection{Outer disk}

As discussed in Paper II, ISM dust mixtures (with small $a_{max}$) cannot
explain the far-IR and mm-wave fluxes of T Tauri stars. 
The similarity between the median Taurus SED and that of TW Hya at wavelengths 
$\lambda \gtrsim 100 \, \mu$m suggests disk models of the type explored in Paper III,
in which we allow for growth to large particles, can in principle
explain the observations.  

Figure 2 shows results for flared, irradiated disk models calculated with the methods of Paper III.
The model SEDs shown are for
$a_{max}$ = 1 mm (which fits the median SED in Taurus;
Paper III), 1 cm,  and 10 cm, calculated
with abundances in the dust mixture usually assumed for protoplanetary
disks (P94), which yield a dust-to-gas ratio of 0.013.
It can be seen that models where grains have grown to $a_{max} \sim 1-10$~cm provide a
much better fit to the long-wavelength SED than the $a_{max} = 1$~mm model,
requiring no or very little additional emission from a wind or non-thermal sources.
Models with $a_{max} \ll 1$~mm fail to reproduce either the sub-mm and mm spectral slopes
or the total flux levels for reasonable disk masses.

Assuming an outer disk radius of $\sim 140$~AU, comparable to the
radius at which Krist \etal (2000) find a rapid decline in disk
density, the disk mass is
$\sim 0.03$, 0.06, and $0.11 \msun$ for the $a_{max} = 1$~mm, 1 cm, and 10 cm
models, or
$\alpha = 5 \times 10^{-4}, 3 \times 10^{-4}$,
and $ 1 \times 10^{-3}$, respectively. The $a_{max} = 1$~cm model has a 
Toomre parameter $Q \sim 1$ at its outer edge and so is near the limit expected for
gravitational stability.  The high mass values are due to the opacity at 7 mm,
which decreases as $a_{max}^{-1/2}$
(for $p = 3.5$, see Paper III), so higher masses are needed to account for the
flux for larger grain mixtures.
For example, for the $a_{max}$ = 10 cm
mixture, the opacity at 7 mm is a factor of 3 lower
than the frequently assumed law $\kappa_{BS} = 0.1 (\lambda/250 \ \mu m)^{-1}$
(Beckwith \& Sargent 1991) and the
slope is slightly flatter, $ \propto \lambda^{0.8}$
(cf. Paper III, Figure 2).
 
Our total disk mass estimates, using the dust-to-gas ratio of 0.013, 
are much higher than the
gas mass obtained by Kastner et al. (1997)
from $^{12}$CO emission, $3.5 \times 10^{-5} \msun$.
This discrepancy may be attributed to
a large degree of molecular depletion
or to the fact that optical depth effects may not have
been properly included (Beckwith \& Sargent 1993);
it could also be due to molecules existing in
the gas phase only in the hot upper atmospheric
layers of the disk where only a small amount of mass resides
(Willacy \& Langer 2000).
An alternative to high disk masses,
not considered in this work, is to have larger
opacities at 7 mm than we obtain; porous aggregates,
specially for amorphous carbon particles, may result
in such larger opacities (Stognienko, Henning \& Ossenkopf 1995).
Our results for grain growth are not unique.  However, within
the assumptions we have made concerning dust opacities, we cannot reproduce
the mm-wave fluxes and spectral slope without including dust particles much
larger than those of a ``standard ISM'' mixture (see Paper III).
Beckwith \& Sargent (1991) noted that optical depth effects could make
the mm-wave spectral slope flatter, and thus reduce or eliminate the
need for grain growth.  However, we are unable to make TW Hya disk sufficiently
optically thick over its large radial extent.

\subsection{Edge of the outer disk}

The well-mixed grain-growth models of Figure 2,
with self-consistently calculated temperature structures,
exhibit too
little emission in the $20-60 \mu$m wavelength region in
comparison with observations.  In principle, disk models
with more small particles could have higher fluxes.   
However, it is apparent that these models also predict far too
much flux at wavelengths $\lesssim 10 \mu$m, because they
are too optically thick in the inner regions;
this suggest that some
clearing has occurred in the inner disk.
Moreover, it suggest that the outer disk should have an optically-thick edge
in which extra heating could be important.

In the case of irradiation of an optically
thick disk with a smoothly-varying thickness,
the stellar flux captured by the disk (and thus the disk
heating per unit area) depends on the cosine of the
angle between the direction of incidence of the stellar beam and the
normal to the disk surface $\mu_0$ (Kenyon \& Hartmann 1987).
In general, stellar radiation penetrates the disk
very obliquely, so  $\mu_0$ is
fairly small, $\sim 10^{-3} - 10 ^{-2}$ (cf. D'Alessio 1996).
However, if the disk had an inner edge, this 
portion of the disk would be illuminated by the star more directly,
increasing $\mu_0$ dramatically and thus increasing the amount of irradiation heating.
We propose that the outer regions of the TW Hya disk
can be described as in \S 4.1, but that it
is truncated at a few AU by a steep,
optically thick region, where most of the mid-IR flux excess
arises (Figure 3). Inside this region lies the inner optically thin disk, which
produces negligible continuum flux. 

To calculate the emission of the disk edge, we have assumed that its 
irradiation surface, that is the surface
where most of the stellar energy is deposited, 
has a shape given by
$z_s = z_o(R_o) exp [ (R - R_o)/ \Delta R]$, where
$z_o$ is the height of irradiation surface
of the outer disk at radius $R_o$
and $\Delta R$ is a characteristic width of the edge.
The edge continuum emission is produced 
in the photosphere, which for simplicity we take as the
irradiation surface; above this layer, there is a hotter optically thin region,
the atmosphere, which we take as isothermal (cf. Chiang \& Goldreich 1997).
We think that a more detailed treatment of the structure of this region
is not necessary given the exploratory nature of this study.
The temperature of the edge photosphere, assuming that 
half of the intercepted flux reaches the
photosphere, is given by
\be
T_{phot}(R) \approx T_* \biggl ( {R_* \over R} \biggr )^{1/2}
\biggl ({\mu_0 \over 2} \biggr )  ^{1/4}.
\en
where $\mu_0 = cos(\theta_o)$ is obtained from 
$\theta_o = \pi/2 - tan^{-1}(dz_o/dR)$.
The flux is evaluated as
\be
 F_{\nu} = \int _{R_{th}}^{R_o} I_{\nu} 2 \pi R dR,
\label{flujo}
\en
where $R_{th} $ is a the radius where the disk becomes optically
thin, and $ I_{\nu} = B_{\nu} (T_{phot}) $. 

The location of the outer disk edge and its width,
characterized by $R_o$ and $\sim  \Delta R$, are fairly
well restricted by the mid-infrared excess,
although the actual shape is not so well constrained
as long as most of it faces the star.
In general we find $\Delta R \sim$ 0.5 AU,
because the shape of the excess is fairly
narrow and thus cannot be produced by a 
region with a large range of temperatures.
In addition, if
$ R_o$ is much smaller than  $\sim $4 AU, then the edge is too hot
and there is too much excess below 10 $\mu$m.
Similarly, if $R_o$ is much larger, there is too little
flux. With $R_o$ $\sim$ 4 AU and $\Delta R$ $\sim$ 0.5 AU,
$\mu_o$ varies from $\sim 0.2$ at $R$ = 3 AU to
$\sim 0.7$ at $R$ = 4 AU, and $T_{phot}$
varies from $\sim $ 80 to 100 in this range.
The resulting continuum flux is shown in 
detail in the lower panel of Figure 4, while
the upper panel shows the fit to the SED of the
composite disk model.

The temperature of the optically thin atmosphere of the edge
is given by 
\be
W(R) \kappa_P^* T_*^4 = \kappa_P(T_{up}) T_{up}^4
\label{tupp}
\en
where $W(R)$ is the geometrical dilution factor, $W(R) = \Omega_*/4 \pi \approx
(R_*/2 R)^2$, for $R >> R_*$, and $\kappa_P^*$ and $\kappa_P$ are
Planck mean opacities calculated at the stellar and local radiation
fields, respectively (cf. Paper I, Paper II, Paper II).
The contribution to the flux from this region is given by
eq. (\ref{flujo}), with $ I_{\nu} = B_{\nu} (T_{up}) \tau_{\nu}$ and
$\tau_{\nu} \approx \kappa_{\nu} \tau_*/\chi_*$, 
where $\tau_*$ and $\chi_*$ are the optical depth and 
extinction coefficient at the characteristic
wavelength of the stellar radiation. 
We follow Natta et al. (2000) taking  $\tau_* \approx \mu_0$,
but in this work we include the effect of scattering
at the wavelengths at which the stellar radiation
is absorbed. The neglect of scattering
in the calculation of $\chi_*$ leads to artificially
large values of $\tau_{\nu}$ and thus of $I_{\nu}$, and
would be only appropriate if scattering was completely forward.
However,  the  asymmetry parameter $g = \prom{\cos \Theta}$,
with $\Theta$ the scattering angle,
is in general $< 1$, and this approximation is not
valid. We have included the effect of scattering using
$\chi_* = \kappa_*+(1-g_*) \sigma_*$,
where $\kappa_*$, $\sigma_*$, are the absorption
and scattering coefficients and $g_*$ the asymmetry parameter
for the assumed dust mixture, all evaluated at the
characteristic wavelength of the stellar
radiation.

We assume that the atmosphere of the edge has small
grains so it can produce emission in the silicate
feature.
We considered different glassy and crystalline
pyroxenes and olivines (with optical properties from  Laor \& Draine 1993,
J\"ager et al. 1994, and Dorschner et al. 1995). We find that
glassy pyroxene $Mg^{0.5} Fe^{0.43} Ca^{0.03} Al^{0.04} SiO_3$
has one of the  highest ratios  $\kappa_{\nu}/\chi_*$
as required for producing a strong band  (see also
Natta et al. 2000). However, 
even with the increasing heating at the disk edge,
the conspicuous silicate emission
feature seen in the SED of TW Hya cannot be explained by emission
from the atmospheric layers of the edge.
This is due in part to the inclusion of
scattering of stellar light in the calculation of
$\chi_*$, as discussed above, and in part to
the adopted dust mixture; organics dominate the
opacity at the stellar wavelength (Paper III, Figure 2), yielding
$\kappa_{10\mu {\rm m}}/\chi_* \sim 1$.
Note that the atmosphere of the optically thick outer disk,
which is irradiated less frontally by the star and
is thus cooler than the atmosphere
of the edge, cannot produce significant silicate
emission either.

\subsection{Inner disk}

Since the region inside the outer disk
edge is not empty because material is still accreting
onto the star (Muzerolle et al. 2000), we have explored the possibility that
a small amount of particles coexists with the accreting gas of the inner disk,
giving rise to silicate
feature emission, while on the other hand still resulting
in negligible continuum flux.
 
We assume that the temperature
of the dust in this region is given by equation (\ref{tupp}). 
The emergent intensity
is calculated as $ I_{\nu} = B_{\nu} \kappa_{\nu} \tau_{10}/ \kappa (10 \mu$m),
where $\tau_{10}$, the optical depth at $10 \mu$m, is a parameter.
This region extends to $\sim 0.02$ AU, the radius where
grains sublimate.
We find that we can fit the silicate feature
with $\tau_{10} \sim 0.05$, see Figure 4, lower panel; this optical depth is small enough
for the disk continuum emission in the near infrared to be negligible.
The best fit to the  profile is achieved with
glassy pyroxene, in agreement with the fact that 
the profile is closer to that in young stars than 
in comets (Sitko et al. 2000).
The grain sizes are in the range $a_{min} \sim 0.9 \mu$m and
$a_{max} \sim 2 \mu$m. Smaller grains have much
higher temperatures and produce too much
emission at short wavelengths, resulting in
a narrower silicate profile than observed.
Bigger grains produce too little emission.

From $\tau_{10} \sim 0.05$, we get a column density 
for the $\sim 1\mu$m dust of $\Sigma_d \sim 4 \times 10^{-3}
{\rm g \, cm^{-2}}$, which implies a mass inside 4 AU of
$\sim 4 \times 10^{25}$ g $\sim$ 0.5 lunar masses.
A strong lower limit to the mass of {\it gas}
in the inner disk can be obtained
from the observed rate of gas accretion onto the star
assuming that the material in the inner disk is in free-fall towards
the star, $M_{gas} >> \mdot R^{3/2}/ (2 G M_*)^{1/2}$.
Inside 4 AU, we obtain $M_{gas} >> 6  \times 10^{-10} \msun$.
Another lower limit can be obtained from $\Sigma_d$,
assuming a normal dust-to-gas ratio, yielding
$\Sigma_g \sim 0.4 \, {\rm g \, cm^{-2}}$
and a mass $M_{gas} > 2 \times 10^{-6} \msun \sim 0.6$ earth masses. If solid material
is hidden into large bodies, then the mass in gas could be
much higher than this value. From this limit,
nonetheless, we
can get an upper limit for the radial velocity of the gas
at 1 AU,
$v_R < \mdot / 2 \pi R \Sigma_g \sim 0.008 \kms
<< v_K (1 AU) \sim 23 \kms$, so the gas probably drifts inwards slowly,
following nearly Keplerian orbits in the inner disk.

Weintraub, Kastner, and Bary (2000) have detected
a flux of $\sim 1.0 \times 10^{-15} {\rm erg \, s^{-1} \, cm^{-2}}$
in the 1-0 S(1) line of ${\rm H_2}$ towards TW Hya.
Using Tin\'e et al. (1997) models,
we can estimate a flux in this line at 55 pc,
$ F({\rm H_2}) \sim 2 \times 10^{-15} ( \epsilon / 10^{-22} {\rm s^{-1}} )
(M_{\rm H_2} / 10^{-8} \msun) {\rm erg \, s^{-1} \, cm^{-2}}$,
where $\epsilon$ is the emissivity per ${\rm H_2}$ molecule.
For a gas temperature of $\sim$ 1000K and a ionization
rate of $3 \times 10^{-10} {\rm s^{-1}}$, estimated with
typical parameters and an X-ray luminosity
$L_X \sim 10^{30} {\rm erg cm^{-3}}$ (Kastner et al. 1997)
using the Glassgold, Najita, \& Igea (1997) corrected
expression, $\epsilon$ is
$2.3 \times 10^{-22} {\rm erg \, s^{-1}}$ for
densities $n_H \ge 10^7 {\rm cm^{-3}}$
(because the line becomes thermalized; S. Lepp,
personal communication).
Assuming a disk height of 0.05 AU, we estimate
$n_H >> 4 \times 10^7 {\rm cm^{-3}}$ inside 4 AU,
so from the mass limit estimated above, we see
that the observed  ${\rm H_2}$ very likely arises
in the inner disk. Firmer conclusions
require a determination
of the gas temperature in the inner disk, 
which is left for future work.

\subsection{Model tests: comparison with VLA 7~mm Data}

In principle, observations of disk surface brightness
distributions can be used to test our physically
self-consistent models based on SED fitting.  The
optical and near-infrared scattered light distributions presented
by Krist \etal (2000) and Trilling \etal (2001) do
not have sufficient resolution to probe our inner
disk region; however, they do provide constraints on
our outer disk structure.  
We find that the scattered light fluxes predicted by
the $a_{max} = 1$~cm model presented above 
matches the observations of Krist \etal (2000) reasonably
well, although Krist \etal find evidence for structure
that cannot be reproduced in detail
with disk properties smoothly-varying with radius, as assumed here.
This work will be reported in a forthcoming paper
(D'Alessio \etal 2001).  

The high angular 
resolution data at 7~mm from the Very Large Array (Wilner et al. 2000)
begin to approach the resolution needed to probe the inner
disk, as well as constrain the thermal dust emission of
the outer disk instead of simply the scattering surface,
and thus provide an additional test of our model.  Moreover,
the imaging data provide an independent test, since the
model was constructed only to match the SED.
The value of $\tau_{10}$ sets an upper limit to
the opacity at $\lambda > 10 \mu$m,
so the inner disk should not contribute significantly to the
emission at 7 mm. We have calculated the intensity
profile of the disk, assuming different values for
the brightness temperature $T_b$ of the edge between
3 and 4 AU, since we do not know the properties
of its interior. An upper limit for $T_b$ is the
surface temperature, $\sim$ 100K, but it could
be lower if it was optically thin at 7 mm,
so we have varied $T_b$ between 60K and 100K.

Figure~\ref{fig:wilner}a  shows the 7~mm images at two angular
resolutions. The ``low'' resolution image ($\sim0\farcs6$, about 35~AU)
emphasizes extended low brightness emission, and the ``high'' resolution
($\sim0\farcs1$, about 6~AU) shows the smallest size scales where
emission remains detectable with the available sensitivity.
Figure~\ref{fig:wilner}b shows the result of imaging the model brightness
distribution at these two angular resolutions using the same visibility
sampling as the observations. 
For a more quantitative comparison, Figure~\ref{fig:wilner}c shows the
residual images obtained by subtracting the model visibilities from the
7~mm data, and then imaging the difference with the standard algorithms.

The low resolution model image, with
the centrally peaked dust emission ($T_b \ge$ 60K), compares
very favorably with the 7~mm images of Figure~\ref{fig:wilner}a;
unfortunately, the higher resolution is still not adequate to resolve
directly the presence of the inner hole.
But in both cases, the residual images show only noise.
The agreement between the model and observations confirms the
insightful parametric modeling of Wilner et al. (2000), who fitted the
observations with radial power laws for the temperature and surface
density. This was to be expected, since Wilner et al. adopted powers
that were consistent with the predictions of irradiated accretion disk
models at large radii.  However, since our models are constructed from
first principles, by solving the disk structure equations in the
presence of stellar irradiation subject to the constraint of steady
accretion, we are able to predict and confirm not only the radial
dependences of the physical quantities (which are not exactly  power
laws), but also their absolute values.  The free parameters of our
model are essentially the disk mass (or equivalently, the parameter
$\alpha$, see \S 3) and the dust mixture. The models were constructed
primarily to fit the SED, but they can also fit the radial brightness
distribution, as the good agreement between the observations and the
model indicate.  On the other hand, although there is no guarantee that the
outer disk is completely steady, or has a constant $\alpha$, or ours
is the correct prescription for the dust mixture, these assumptions
seem to provide a good description of the physical situation in the
disk of TW Hya.

\section{Discussion} 

The lack of infrared flux excess at wavelengths below 10 microns
could be easily understood if the inner disk was evacuated inside a
few AU. However, the fact that material is still being transfered onto
the star, as evidenced by the broad emission line profiles and the
ultraviolet excess emission (Muzerolle et al. 2000), indicates that
gas still exists in the inner disk.  However, this material must be
optically thin to explain the flux deficit in the near infrared.  If
the outer disk extended inwards to the magnetospheric radius, then the
surface density $\Sigma$ at 2 AU would be $\sim 320 {\rm g \,
cm^{-2}}$, implying a surface density in solids of $\Sigma_{dust} \sim
3 {\rm g \, cm^{-2}}$ (for a dust to gas mass ratio of 0.01).  To
obtain an optical depth at the near-infrared of $< 0.05$ with this
amount of dust, the opacity should be of order $\kappa_{dust} < 0.05 /
\Sigma_{dust} \sim 0.02 {\rm cm^2 g^{-1}}$.  If we assume that the
grains are large enough that the opacity per gram of grain is $\sim
Q_{ext} \pi a^2 / \, (4/3) \pi \rho_g a^3$, with $Q_{ext}$ the
extinction cross section and $\rho_g$ the grain density, both or order
1, then $\kappa_{dust} \sim 1/a$. Using this estimate, we find that
the original solids should have grown to sizes $ a > 50$ cm to account
for the low optical depth in the near infrared.  While this estimate
is very rough, it suggests that very significant grain growth has
probably already taken place in the inner disk of TW Hya, especially
if the solid to gas ratio has increased significantly from solid
matter left behind as gas flows onto the star.  This inside-out
coagulation and settling is consistent with the evolution of solid
particles predicted in solar nebula theories.  Dust grains are
expected to coagulate by sticking collisions at a rate $\propto 1 /
\Omega^2 \propto P^2$ and reach a maximum size at the midplane in a
few $\times 1000 P$, where $\Omega$ is the Keplerian angular velocity
and P the orbital period (Safronov 1972; Goldreich \& Ward 1973;
Weidenschilling 1997; Nakagawa 1981). Since $ P \sim 1.4 \times 10^3
(R/100 AU)^{3/2}$ yrs, coagulation and settling are expected to occur
first in the inner disk.

In the outer disk, a dust mixture where grains have reached $\sim$ 1 cm
and dust and gas are well mixed can explain the observations,
within the framework of our assumptions.
This interpretation is consistent with the Weidenschilling's (2000)
argument that grain growth to at least 1 cm is necessary
before particles start settling to the midplane,
because smaller grains get stirred around by turbulence,
and we suggest that turbulence plays an important role
in the evolution of dust in protoplanetary disks.
Grains in the upper layers can absorb stellar radiation
and 
efficiently heat the disk and make it flare,  
resulting in the large far infrared 
and millimeter fluxes observed.
Nonetheless, grains in the outer disk have grown
significantly
from $\sim 1 $ mm, the size that characterizes the median SED of Taurus (Paper III).
To get high mm fluxes for such large
grains, the disk mass has to be high, $\sim 0.06 \msun$, since
the opacity at 7 mm decreases as grains grow (cf. Paper III, \S2). 
This high value for the mass is surprising; it is similar or
higher than values typically found
for disks around 1 Myr old T Tauri stars (Beckwith et al. 1990), and comparable
to that of objects emerging from the embedded phase
(D'Alessio, Calvet, \& Hartmann 1997).
On the other hand, 
as we discussed in Paper III, disk masses
may have been underestimated if the standard power law opacity
is applied indiscriminately to disks where
significant grain growth may have taken place.

It may be possible that the 3.6 cm data point
contains a significant contribution from hot (ionized) stellar plasma.
This contribution is unlikely to come from
a wind. If the flux emitted by the 
wind is proportional to the mass loss rate, then
scaling down from the wind/jet contribution
in the case of HL Tau, which has $\mdot \sim 10^{-6} \msunyr$
to the case of TW Hya, and assuming a ratio of mass
loss rate to mass accretion rate of $\sim 0.1$ (Calvet 1998),
we find that this contribution at 3.6 cm should be $\sim 2 \times 10^{-4}$
mJy, too low to be important. A non-thermal contribution due to stellar surface
activity is still possible, given the high X-ray
flux of TW Hya (Kastner et al. 1997). Second epoch observations
should resolve this question, 
because any non-thermal contribution should be highly time-dependent.
In any event, the high flux at 7 mm, at which a non-thermal
plasma contribution is much less likely, indicates
that the grain mixture must contain grains that
have grown at least to $\sim$ 1 cm, otherwise
the opacity would already turn to the optical
limit and drop, as is apparent in Figure
2 (cf. Paper III, Figure 2). 

The peculiarities of the disk of TW Hya suggest an advanced
state of evolution of the dust. In addition, the
presence of a steep transition between the inner
and outer disk suggest the action of an agent other than 
dust settling to the midplane, which in first
approximation would be expected to 
produce effects that vary monotonically
with radius. We speculate that we
may be seeing the outer edge of a gap opened
by the tidal action of a growing protoplanet.
Numerical models of gap formation in disks result in 
a sharp drop of the surface density at the gap and
enhancements near the edges, as disk
material is pulled away from the protoplanet
and into the outer disk (Bryden et al. 2000;
Nelson et al. 2000). The edge of the gap would be
facing the star, and if it was optically thick,
would result in excess emission similar to that modeled schematically
by our truncation region. 
The situation could be comparable to the one considered
by Syer \& Clarke (1995), who studied the evolution
of the disk spectrum under the presence of a
planet formed in the disk, in the case in which the mass of the
protoplanet was larger than the mass of the disk at the planet location.
In this case, the disk cannot push the planet
inwards at the local viscous time-scale
and material forms a reservoir upstream of the gap
(Lin \& Papaloizou 1986). The surface density enhancement
at the edge could make it optically thick,
so the midplane temperature would not increase
as much as the surface temperature; the scale
height, in this case, would not increase significantly,
justifying
our assumption of similar heights for the edge region 
and outer disk and the neglect of the effects of shadowing.
The details of the structure of this region, in any event, are of necessity
very schematic. The extrapolated
mass of the disk inside $\sim$ 4 AU is
$\sim 10^{-3} \msun \sim$ 1 Jupiter mass, so
this mechanism would work if a giant planet were forming 
in this region of the disk.

Finally, we address the question of what TW Hya tells us about protoplanetary
disk evolution.  Although we do not have direct measurements of
the gas mass of the outer disk, it would be very surprising if the gas were
strongly depleted there, as suggested by
Zuckerman et al. (1995).  Both the SED and the scattered light emission
as a function of radius follow the predictions of models in which 
the dust is well-suspended in the gas, producing finite scale heights
of the dust disk.  Based on this inference, we suggest that the existence
of the (fairly massive) TW Hya disk is evidence that neither UV radiation
nor stellar winds can efficiently remove gas at radii $\gtrsim 5$~AU
over timescales of 10 Myr for stars formed in dispersed, isolated environments.

The survival of disk material at 10 Myr also places constraints on viscous timescales.
The surface density of an accretion disk evolves
with time, as viscous dissipation makes the disk
transfer mass onto the star and expand to conserve angular
momentum in a viscous time-scale given by $R_c^2/ \nu$,
where $R_c$ is a characteristic radius and $\nu$
the viscosity (cf. Pringle 1981). 
Here we use the similarity solutions for disk evolution of Hartmann et al. (1998),
which assume temperature and surface density distributions comparable to our outer
disk models.  The viscous time-scale in these solutions is
$t_s \sim R_1^2 / \nu \sim 8 \times 10^4 ( 10^{-2} /\alpha )
(R_1 / 10 AU)$ yrs, where $R_1$ is the radius that contains $\sim$ 60 \%
of the initial mass (at $t = 0$), and
the viscosity $\nu$ is expressed in terms of 
the standard Shakura-Sunyaev (1973) $\alpha$ parameterization.

Adopting $\alpha \sim 5 \times 10^{-4}$ and $R_1 \sim 100$ AU,
then $t_s \sim 1.6 \times 10^7$ yrs.  In the similarity model, 
the mass of the disk falls in time as $M_d \propto T^{-1/2}$,
where $T = 1 + t/ts$.  Thus, if the total disk mass at 10 Myr is $0.06 \msun$,
then the initial disk mass must have been $0.08 \msun$, comparable to that of objects
still surrounded by infalling envelopes, like HL Tau (D'Alessio, Calvet, \& Hartmann 1997). 
The surface density drops below a $\Sigma \propto R^{-1}$ power-law by a factor of
$1/e$ at a distance $R_1(T) = R_1 T \sim 160$~ AU, close to 
the $\sim 140$~AU transition in midplane density found in Krist \etal (2000).
With these values of disk mass and radius, the value of the
Toomre parameter $Q$ remains close to
unity and so gravitational instabilities (which could transfer angular momentum
rapidly and thus invalidate the purely viscous 
evolution model) need not arise.

The mass accretion rate for this constant-$\alpha$ model 
\footnote{Note that
a factor ($M_d(0)/0.1 \msun$) is missing in
expression (39) of Hartmann et al. (1998).}
at an age of 1 Myr
is $\mdot = M_d(0)/(2 t_s T^{3/2}) \sim 2 \times 10^{-9} \msunyr$,
near the lowest values observed in accreting T Tauri stars of this age 
(Gullbring et al. 1998; Hartmann et al. 1998).
The mass accretion rate at 10 Myr is then $\sim 1 \times 10^{-9} \msunyr$, 
which is slightly higher than estimated by Muzerolle \etal (2000).  It would not be
surprising if
accretion is being halted in the inner disk, and that inner and outer disk are becoming
decoupled.
If material is prevented from reaching the inner disk, the material
inside the gap would drain onto the central star on a viscous time
at $\sim$ 4AU, which with $\alpha \sim 10^{-3}$,
would be $\sim 3 \times 10^5$ yrs.  Such a short timescale suggests that
the probability of observing a system in this stage is very low, but still possible.

Instead of the $\alpha \sim 10^{-3}$ adopted above,
Hartmann \etal (1998) estimated a typical value of $\alpha = 10^{-2}$ for T Tauri
disks based on mm-wave disk mass estimates and assuming that these 
disks expand to $\sim 100$~AU in $\sim 1$~Myr.  If the TW Hya disk 
(the region containing most of the mass) is not much larger
than 200 AU in radius, then $\alpha$ cannot be this large.  A smaller viscosity
could be consistent with the typical T Tauri data if disk masses are a few
times larger than typically estimated, as suggested by D'Alessio \etal (2000).
 
One remaining question is why TW Hya's disk evolution has been so much slower
than that of other members of the association with presumably similar ages.
In part, this must be due to the presence of binary (or multiple) stellar
companions in several other systems (e.g., HD 98800; Soderblom \etal 1998)
which can disrupt disks.
In addition, we conjecture that TW Hya may simply have had a larger initial
disk radius (equivalent to a large $R_1$ in terms of the similarity solution
above), and thus a lower average surface density, than other systems.
A larger radius would lead to slower viscous evolution for a given $\alpha$.
Also, in contrast to timescales for dust coagulation and sedimentation, which
depend mostly on orbital period (see above), 
theoretical models of the runaway growth of giant planets indicate that
the timescale for such growth is a very sensitive function of surface
density (Pollack et at. 1996).
The lower surface density of the disk around TW Hya
would imply a slower growth of giant planets. This would imply,
in turn, that the final clearing of disks
is due to sweeping-up of material by large bodies. 

\section{Conclusions}

The disk of TW Hya seems to be in advanced state of dust
evolution.  A planet or other large perturbing body 
may have already formed, opening
a gap with outer edge around $\sim$ 4 AU, and 
although there is still material in
the inner disk, it is rapidly dissipating. 
In the outer disk, grains may be reaching the size
necessary to start settling towards the midplane.
This interpretation can consistently explain the
large degree of activity of TW Hya, the lack
of disk emission at near-IR wavelengths and
the large fluxes beyond $\sim$ 10 $\mu$m,
as well as the 7 mm images. 

Evacuated regions in the disks
are also known to result from the effects of companion stars (GG Tau),
and in some cases disk accretion can still occur, especially if the binary
has an eccentric orbit (DQ Tau).  Indeed, HD98800, another member of the TW
Hya association, shows evidence for an inner disk hole; because this system
is quadruple, it is quite likely that the inner disk regions have been evacuated
by the companion stars. Although we cannot rule out such a possibility in TW Hya,
the fact that accretion still occurs onto the central star, which is not the
case in HD 98800, suggests that the companion body responsible for the gap
is much less massive than a typical companion star, and thus much less effective
in clearing out the disk.  Interferometric imaging should be attempted to constrain
the mass of any companion object, expected to have a separation of order
2-3 AU (0.04 - 0.06 arcsec at 55 pc).

Acknowledgments. We gratefully acknowledge discussions
with Edwin Bergin and Stephen Lepp. This work has
been supported by NASA Origins of Solar Systems grants NAG5-9670
and  NAG5-9475.
P.D. has been supported by Conacyt grant J27748E and
a fellowship from DGAPA-UNAM, M\'exico.

\clearpage
\begin{figure}
\plotone {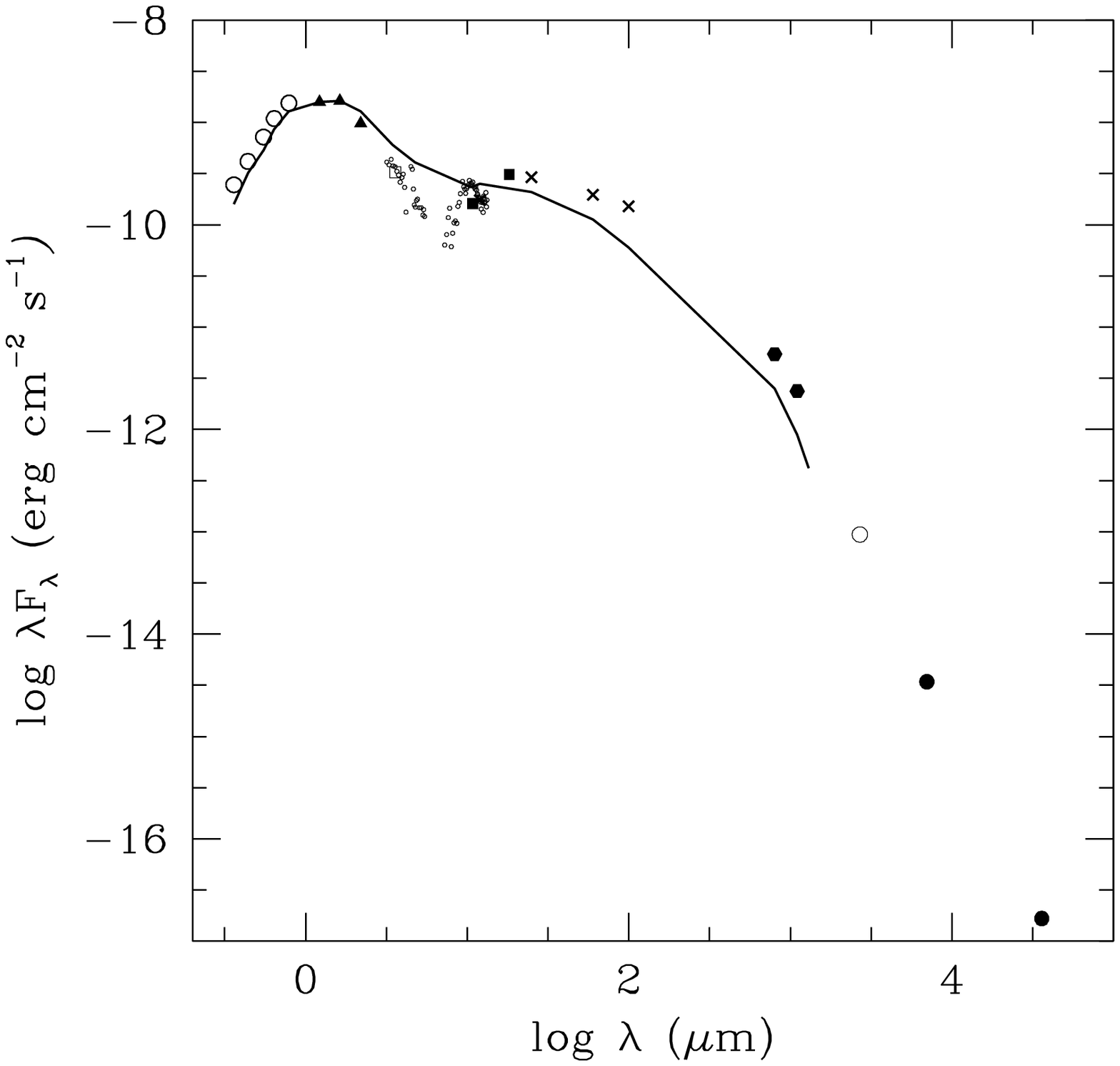}
\caption{Spectral energy distribution of TW Hya.
Observations are from 
Rucinski \& Krautter (1983; average of
UBVRI measurements) (open circles),
Webb et al. (1999) (triangles),
Sitko et al. (2000) (small open circles),
this paper (open square),
Jayawardhana et al. (1999) (squares),
IRAS (de la Reza et al. 1989; Gregorio-Hatem et al. 1987; crosses),
Weintraub et al. (1989) (hexagons), 
Wilner (2001) (open circle),
and
Wilner et al. (2000) (filled circles)
The solid line is the median SED of classical
T Tauri stars in Taurus (Paper II). 
Note the excess flux at mid-infrared
wavelengths and the flux deficit below 10 $\mu$m (see text).
}
\end{figure}

\begin{figure}
\plotone{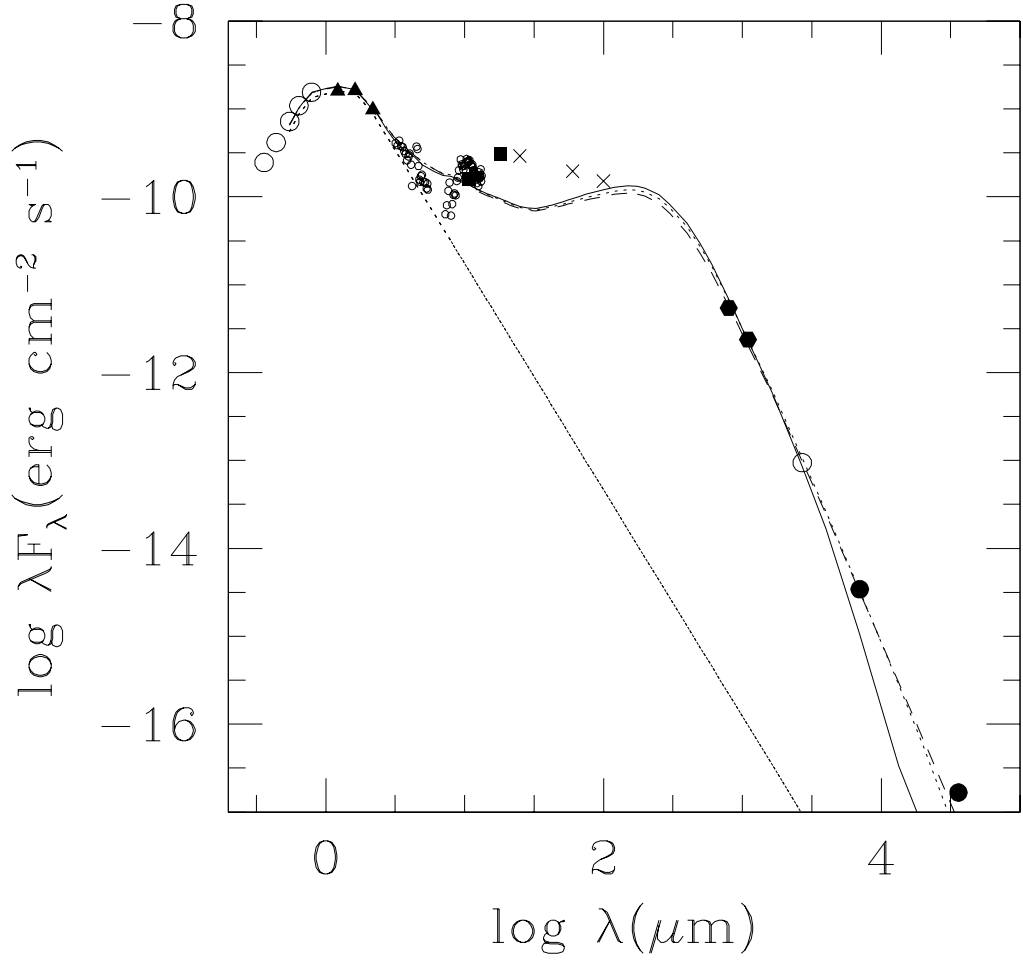}
\caption{SED of TW Hya and disk models with gas and dust well
mixed and extending to the magnetospheric
radius, for $a_{max}$ = 1 mm (solid line),
1 cm (dotted line), and
10 cm (long dashed line).  Larger maximum grain sizes produce a better match
to the sub-mm and mm-wave spectrum.
The SED of the
stellar photosphere is also shown for comparison (short dashed line.)
}
\end{figure}

\begin{figure} 
\plotone{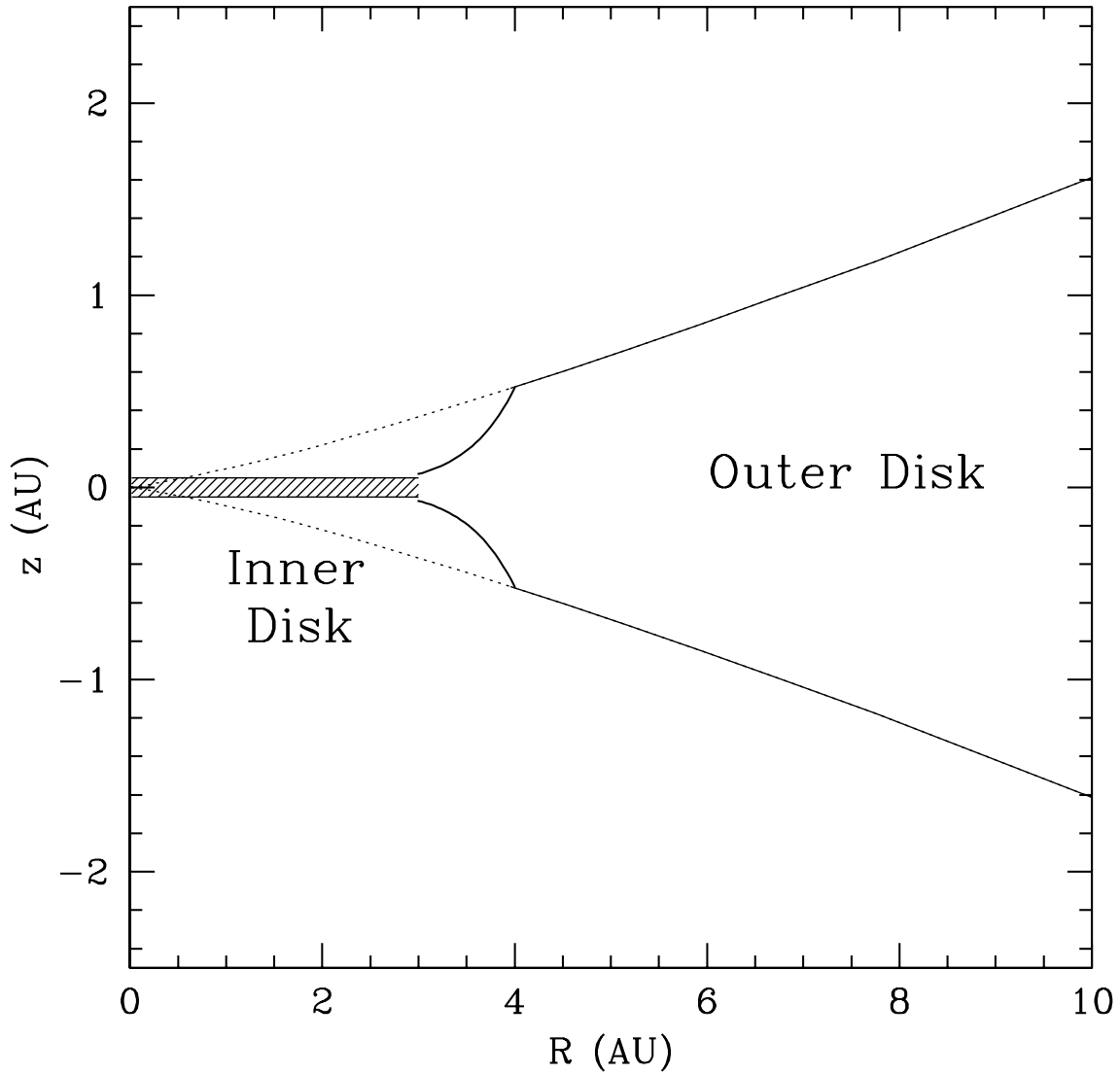}
\caption{Model disk adopted for TW Hya. The outer disk,
where grains have grown to $\sim$ 1 cm,  has a
edge at $R \sim 3-4$ AU, and surrounds the inner
optically thin disk, which has a vertical optical depth at $10 \mu$m
$\tau_{10} \sim 0.05$. Gas still exists in the inner disk
accreting onto the star through a magnetosphere. A minute
amount of $\sim$ 1 $\mu$m dust permeates this gas.
The dotted line is the surface of the outer
disk if it extended inward; the resulting SED for this
model is shown in Figure 4.
} 
\end{figure}

\begin{figure}
\plotone{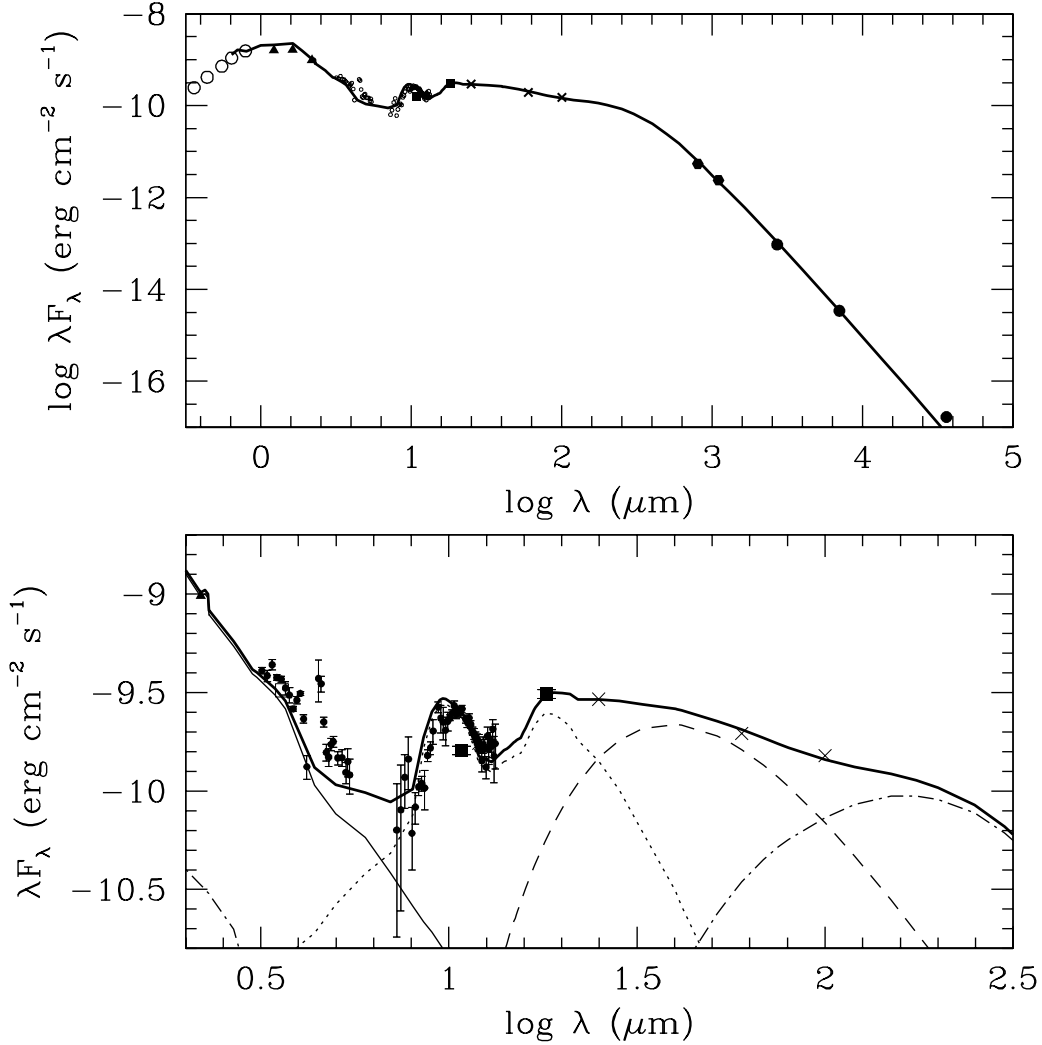}
\caption{
Above: Fit to the TW Hya SED
with our composite disk model. Below:
Detail of the infrared region.
The emission from each region is indicated:
outer disk (dot-dash), edge of outer disk
(dashed), inner disk (dotted), star (light solid),
total (heavy solid). The stellar spectrum
is taken from the Allard \& Hauschildt (1995)
M dwarf library, where a model with appropriate
effective temperature and gravity (log g =4) could
be found. The model has been scaled to the observations
at K (2.2 $\mu$m). The excess near $\sim 4.5 \mu$m could 
be CO fundamental emission; the amount of excess (if any) at
$\lambda < 5 \mu$m is very uncertain, as it depends
critically on the effective temperature (and model) for
the stellar photosphere.
}
\end{figure}

\begin{figure} 
\plotfiddle{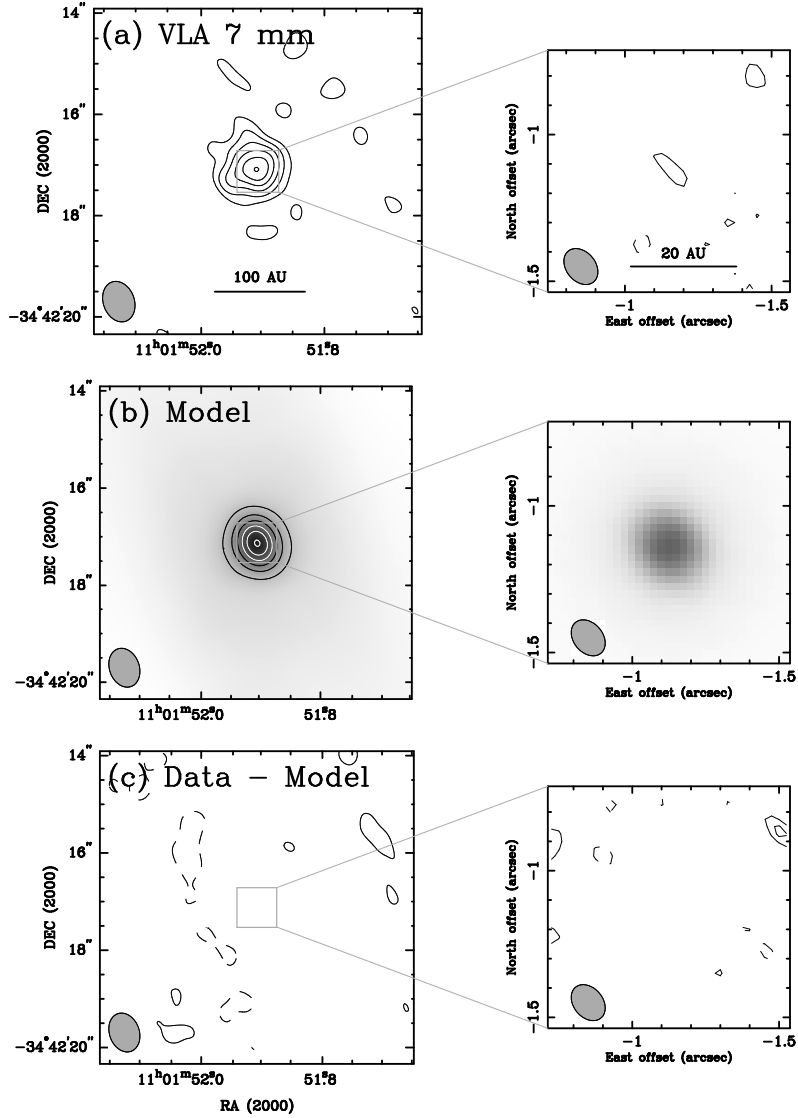}{5.5in}{0}{65.0}{65.0}{-200.0}{-60.0}
\caption{
{\em (a)} VLA 7~mm images of TW Hya at two resolutions,
from Wilner et al. (2000).
The synthesized beam sizes are $0\farcs82\times0\farcs61$ p.a. 20 (left)
and $0\farcs13\times0\farcs10$, p.a. 39 (right). The contour levels
are $\pm(2,3,4,5,6)\times$ the rms noise levels of 0.5 and 0.4 mJy/beam.
Negative contours are dashed.
{\em (b)} Simulated VLA 7~mm images of the disk model brightness
distribution described in the text. The contours and beam sizes are
the same as in {\em a}. The grey scale shows extended emission,
most of which remains undetectable with the available VLA sensitivity.
{\em (c)} Difference images obtained by subtracting visibilities derived
from the model from the VLA 7~mm data.
Again, the contours and beam sizes are the same as in {\em a}.
The residual images show no significant signal.
} 
\label {fig:wilner}
\end{figure}

\end{document}